\documentclass[12pt,preprint]{aastex}

\usepackage{graphicx}
\usepackage[usenames,dvipsnames]{color}

\shorttitle{Polarization survey of the LC / Loop I interface}
\shortauthors{Santos, Corradi \& Reis}

\begin{document}


\title{Optical Polarization Mapping toward the interface between the Local Cavity and Loop I}


\author{F\'abio P. Santos, Wagner Corradi, and Wilson Reis}
\affil{Departamento de F\'isica – ICEx – UFMG, Caixa Postal 702, 30.123-970 \\
Belo Horizote – MG, Brazil}
\email{[fabiops;wag;wilsonr]@fisica.ufmg.br}




\begin{abstract}

The Sun is located inside an extremely low density and quite irregular volume
of the interstellar medium, known as the Local Cavity (LC). It has been widely believed that
some kind of interaction could be occurring between the LC and Loop I, a nearby
superbubble seen in the direction of the Galactic Center.
As a result of such interaction, a wall of neutral and dense material,
surrounded by a ring shaped feature, would be formed at the interaction zone.
Evidence of this structure was previously observed
by analyzing the soft X-ray emission in the direction of Loop I.
Our goal is to investigate the distance of the proposed annular region and map
the geometry of the Galactic magnetic field in these directions.
On that account, we have conducted an optical polarization survey to
$878$ stars from the Hipparcos catalogue.
Our results suggest that the
structure is highly twisted and fragmented, showing very discrepant distances along the
annular region: $\approx 100$ pc to the left side and $250$ pc to the right side, 
independently confirming the indication from a previous photometric analysis.
In addition, the polarization
vectors' orientation pattern along the ring also shows a widely different behavior toward
both sides of the studied feature, running parallel to the ring contour in the left side and
showing no relation to its direction in the right side. Altogether, these evidence suggest a highly
irregular nature, casting some doubt on the existence of a unique large-scale ring-like structure.

\end{abstract}





\section{Introduction}

\hspace{1.0em}
The Sun is located roughly at the center of a large interstellar feature
in the Local spiral arm (the Orion Spur). This conspicuous
structure, known as Local Cavity (LC) or Local Bubble, consists
of a very irregular, low density volume ($n_{HI} < 0.005$ cm$^{-3}$) of the interstellar
medium (ISM), whose borders extend from $\approx 65$ to $250$pc, depending on the
direction to which we observe it \citep{coxreynolds1987,welsh1994,lallement2003,vergely2010}.
This cavity is surrounded by several other interstellar bubbles that are sometimes associated
to strong star-forming activity, and are frequently believed to have
been generated by supernovae (SN) explosions and very intense stellar
winds from massive OB stars \citep{quigley1965,berkhuijsen1971,weaver1979}.

Several efforts have been made to build a tridimensional view of the LC,
mainly by using Na I observations in the direction of nearby stars, which
is a suitable tracer of the neutral gas \citep{welsh1994,sfeir1999,lallement2003,vergely2010}. Information on the shape
and size of this structure can also be inferred from ultraviolet
interstellar absorption lines
\citep{frisch1981,frisch1983,paresce1984,centurion1991,welsh1994,redfield2000,sallmen2008,welsh2005},
interstellar reddening
\citep[][Reis et al. 2010, in preparation]{franco1990,corradi1997,corradi2004,frisch2007,reis_corradi2008,vergely2010}
and interstellar polarization
\citep{tinbergen1982,reiz1998,heiles1998,heiles2000,leroy1999}.
The overall shape of the LC suggests that it is being compressed due to the expansion
of the neighboring bubbles, with a narrower dimension along the Galactic Plane (GP),
and probably opened in the direction of the Galactic halo. Furthermore,
this ``chimney" structure is slightly tilted relative to the GP ($\approx 20^{\circ}$),
and perpendicular to the Gould's Belt, a large complex of young massive OB stars
surrounding the Local ISM \citep{welsh1994,sfeir1999,lallement2003,vergely2010}.

In the direction of the Galactic Center, we find a particularly
interesting structure known as the Loop I superbubble, which is 
centered at the Scorpio-Centaurus OB association (Sco-Cen, $l\approx329^{\circ},b\approx+17.5^{\circ}$), and probably
created due to its intense stellar activity \citep{blaauw1964,degeus1992,sartori2003,preibisch2008}.
This neighboring interstellar bubble, located at $d\approx 130$pc and defined by a large sky-projected diameter
($\sim115^{\circ}$), has been better revealed by radio continuum observations,
which show several arc-shaped shells of interstellar material \citep{berkhuijsen1971,iwan1980,haslam1982,heiles1998}.

The proximity between the Local and Loop I bubbles led some authors to believe
that some kind of interaction could be taking place between them \citep{coxreynolds1987,centurion1991}.
In fact, by analyzing shadowing effects at the wide-angle soft X-ray survey
from ROSAT ($0.25$keV), \citet{egger_aschenbach1995} proposed that the collision between
both bubbles led to the formation of a wall of neutral gas surrounded by a dense
interstellar ring feature at the interaction zone. This conclusion was inspired
by collisional models of spherical shock fronts, which revealed that
a dense interacting wall would arise, encompassed by an even denser annular
feature, if at least
one of the interacting shells have reached the radiative stage before the collision
occurred \citep{yoshioka1990}. This idea was supported by an anticorrelation between the shadows from the
soft X-ray maps and the neutral hydrogen (HI) emission from the local ISM \citep{egger_aschenbach1995,breit2000}.

Up to date, several attempts have been made to determine the distance to the
supposed interacting region, leading to widely different results.
\citet{centurion1991} suggested a distance of $40\pm 25$ pc, from the analysis
of ultraviolet spectra to eight stars at the region defined by
$315^{\circ}<l<330^{\circ}$ and $15^{\circ}<b<25^{\circ}$.
\citet{egger_aschenbach1995} used HI column densities data from \citet{fruscione1994}
to determine a distance of $70$ pc, defined by a jump in $N_{H}$
from $\leq 10^{20}$ cm$^{-2}$ to $\geq 7\times 10^{20}$ cm$^{-2}$ at this distance.

Using $E(b-y)$ color excess data and high resolution spectroscopy
in the direction of the Southern Coalsack, Chamaeleon, and Musca dark clouds
\citet{corradi1997,corradi2004} suggested that the interaction zone
is twisted and folded, located at $120-150$ pc along this line-of-sight.
As previously pointed out by \citet{dame2001}, several dark clouds
($\rho$ Oph, Lupus, R CrA, G317-4, Southern Coalsack, Chamaeleon, and Musca)
are located at the same mean distance of $150$ pc, in the direction of Loop I.
However, \citet{welsh2005} identified the presence of an interstellar cloud
in the direction of $(l,b)\approx(330^{\circ},+18^{\circ})$, at approximately
$90$ pc from the Sun, suggesting that this could be part of the
interface between the Local and Loop I bubbles.

Recently, \citet{reis_corradi2008} used a larger sample of $E(b-y)$ color excess data
distributed along the entire interface region to map the interstellar dust distribution.
The analysis led to the conclusion that the expected transition
from nearly $E(b-y)=0\fm015$ to $E(b-y)\approx 0\fm070-0\fm100$
\citep[which corresponds to the ring's density, as proposed by][]{egger_aschenbach1995},
occurs at the western (left) side at $110\pm 20$pc while the eastern (right side) transition cannot be clearly
defined before $280\pm 50$ pc.

Moreover, the structure of the interstellar magnetic field along the borders
of the LC have been previously studied by several polarimetric surveys
\citep{mathewson_ford1970,tinbergen1982,reiz1998,leroy1999,heiles2000}.
It is known that the local ISM is filled by an irregular, large-scale magnetic flux of average
intensity $\langle B\rangle\approx 2.2\mu$G \citep{heiles1998,beck2001,heiles2005}.
Although no final conclusion has been reached in relation to the dominant physical mechanism responsible
for the alignment of the interstellar dust particles, it is generally accepted that
in the majority of cases, grain alignment occurs with the grain's major axis perpendicular to the magnetic
field direction
\citep{hall1949,hiltner1949,davis_greenstein_1951,jones_spitzer1967,codina1976,purcell1979,lazarian1995a,lazarian1995b,draine_wein_1996,draine_wein_1997,fosalba2002,lazarian2007}. 
This configuration of dust grains gives rise to an anisotropic extinction
which results in a partially polarized transmitted light beam from a distant star,
with position angle in the same direction as the field $\mathbf{\bf{B}}$. Therefore, there
is a strong correlation between the direction of the plane-of-sky projected component of $\mathbf{\bf{B}}$
and the polarization vectors, which can be used as a powerful tool to map the Galactic magnetic field,
as well as to probe the nature of the interstellar dust particles.

All-sky polarization surveys exhibit a large-scale vectors distribution pattern
which is generally correlated to the direction of the local interstellar structures. Specifically,
it is frequently found that polarization vectors may be aligned roughly perpendicular or parallel
to the interstellar filaments, depending on several physical and geometrical factors,
including projection effects \citep{heiles1998,fosalba2002,whittet2003,heiles2005}.
Particularly along the Galactic plane it is noted an overall distribution of polarization
angles which is mainly horizontal (i.e., parallel to the plane). Such trend reflects
the morphology of the local magnetic field, directed mainly along the
local spiral arm. In fact, this predominant orientation along the Galactic plane
vanishes when the local magnetic field is viewed face-on along its ``poles" at $l\approx(80^{\circ},260^{\circ})$,
which is roughly coincident with the direction of the local spiral arm.

Large-scale mappings of the magnetic field structure in other spiral galaxies show
that the orientation parallel to the galactic plane is a general trend which may
be attributed to differential rotation of the galactic disk and magnetic flux
freezing with the interstellar matter
\citep{zweibel1997,beck2002}.

In this work, we present an optical polarimetric survey
in the direction of the interface between the LC and Loop I. Section \ref{obsdata} provides
a description of the observational data, as well as the reduction process.
The results and analysis of the correlation between polarimetric and color excess data,
as well as of the spatial distribution of the polarimetric vectors and
polarization degree as a function of distance are shown on sections \ref{correlation_p_colourexcess}, \ref{polvec} and \ref{poldist}.
Discussion of the results is carried out on section \ref{discussion} and
the conclusions are shown on section \ref{conclusions}.

\section{Observational Data}
\label{obsdata}

\hspace{1.0em} The polarimetric data used in this work were collected
at OPD (Observat\'orio Pico dos Dias, LNA/MCT, Brazil) during $63$ nights
distributed between 2007 August and 2009 February. We have used both the
$60$cm and $1.6$m telescopes, equipped with an imaging polarimeter attached to a specially adapted CCD camera.

The polarimeter consists of a rotating half-wave retarder, a fixed analyzer, and
a filter wheel. The half-wave retarder is allowed to rotate in discrete steps of $22.5^{\circ}$,
so that 16 steps corresponds to a complete rotation ($360^{\circ}$). Its effect
on the incident stellar light is to cause a rotation of the polarization plane of the linearly polarized component.
For instance, if the stellar initial polarization angle (relative to
the North Celestial Pole - NCP) is $\theta$, and $\psi$ is the angle between the retarder's
optical axis and the NCP, the outcome is a light beam with polarization angle
of $2\psi - \theta$ \citep{serkowski1974}.

The analyzer was a Savart plate, i.e., a birefringent double calcite prism which subdivides
the incoming light beam into two orthogonally plane-polarized beams. Thereafter, both
beams pass through a filter, and are
simultaneously detected by the CCD, which registers the individual beams' intensities
at each position of the half-wave retarder
\citep[for a complete description of this instrument, see][]{magalhaes1996}.

Thus, the Q and U Stokes parameters are obtained by fitting a
4-cosine modulation amplitude curve \citep{magalhaes1984,santos2009}, so that the polarization degree ($P$)
and the polarization angle ($\theta$) may be calculated, respectively, by
\begin{center}
$
P=\frac{\sqrt{Q^{2}+U^{2}}}{I} \ \ \ \ \ \ \ \mathrm{and} \ \ \ \ \ \ \ \theta=\frac{1}{2}\arctan{U/Q}
$
\end{center}
\noindent ($I$ represents the intensity of the incident light beam). The simultaneous observation
of both light beams allows a differential calculation of the polarization parameters, i.e.,
independent of the atmospheric variations.

   \begin{figure}[!t]
   \centering
   \includegraphics[width=0.5\textwidth]{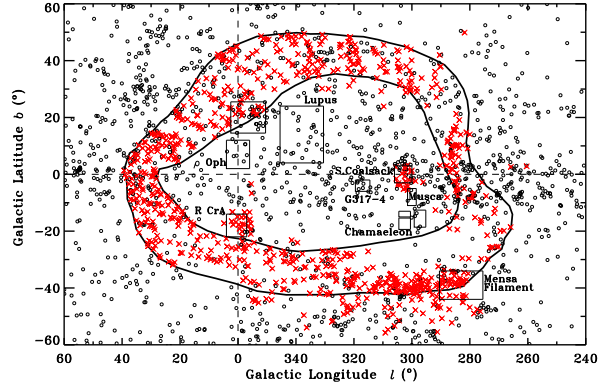}
      \caption{Polarimetric data distribution along the interface between the LC and Loop I:
               crosses ({\color{Red}$\times$}, colored in red at the electronic version) represent objects
               observed at OPD/LNA and open circles ($\circ$)
               represent the data acquired from \citet{heiles2000} catalogue. The thick contour
               indicates the position of the annular feature proposed by \citet{egger_aschenbach1995},
               and the boxes roughly delimit the position of the indicated dark clouds.
               The data collected at OPD are mainly concentrated along the interaction ring.
              }
         \label{distr_data}
   \end{figure}

\begin{table}[!b]
\caption{\label{poldata_short} Polarimetric data obtained at OPD.}
\begin{tabular}{rrrrrr}
\tableline\tableline
HIP& $l(^{\circ})$ & $b(^{\circ})$ & d(pc) & $P_{V}(\%)$ & $\theta_{gal}(^{\circ})$  \\
\hline
88294    & $355.277$ & $-6.689 $ & $ 222$ & $ 0.772\pm 0.043 $ & $ 30.4\pm 1.9  $ \\
90969    & $355.998$ & $-13.331$ & $ 151$ & $ 0.293\pm 0.023 $ & $  6.4\pm 2.5  $ \\
94157    & $355.552$ & $-20.975$ & $ 256$ & $ 0.317\pm 0.026 $ & $ 80.8\pm 2.5  $ \\
105062   & $353.880$ & $-44.005$ & $  42$ & $ 0.050\pm 0.024 $ & $ 28.7\pm 13.8 $ \\
105133   & $350.192$ & $-43.855$ & $ 133$ & $ 0.035\pm 0.010 $ & $ 18.6\pm 8.3  $ \\
105180   & $354.923$ & $-44.319$ & $ 344$ & $ 0.172\pm 0.028 $ & $ 67.2\pm 4.8  $ \\
105391   & $354.142$ & $-44.695$ & $  53$ & $ 0.027\pm 0.053 $ & $ 24.5\pm 56.2 $ \\
105425   & $348.398$ & $-44.217$ & $ 112$ & $ 0.104\pm 0.029 $ & $139.8\pm 8.1  $ \\
105685   & $352.919$ & $-45.208$ & $  97$ & $ 0.015\pm 0.029 $ & $ 91.0\pm 55.4 $ \\
12654    & $297.063$ & $-36.043$ & $  59$ & $ 0.023\pm 0.013 $ & $168.8\pm 16.2 $ \\
\hline
\end{tabular}
\tablecomments{
Table \ref{poldata_short} is published in its entirety in the electronic edition of
the Astrophysical Journal. A portion is shown here for guidance regarding its form and content.
At the columns above, HIP represent the Hipparcos identifier, $(l,b)$ are the Galactic coordinates,
d is the trigonometric distance from the Hipparcos catalogue, $P_{V}$ is the polarization
degree (V filter), and $\theta_{gal}$ is the Galactic polarization angle.
}

\end{table}

Standard stars were obtained from various catalogues
\citep{serkowski_etal_1975,bailey_hough_1982,hsu_breger1982,bastien1988,tapia1988,turnshek1990,schmidt_et_al_1992,gil_benavidez_2003}
and were observed at least twice every night (two polarized and two unpolarized standard stars),
in order to determine the zero-point of the polarization degree and angle.
Moreover, observations of standard stars were performed by using 16 positions of the
half-wave plate, to minimize the uncertainties of the computed polarimetric parameters.
Program stars were observed with the usually adopted 8 positions of the half-wave plate.
Instrumental polarization was mainly negligible when compared to the typical
polarization errors provided by this scheme ($<0.1\%$).

The reduction of the CCD images was performed using the IRAF standard routines from
the CCDRED package. The polarimetric parameters were calculated for each star,
using PCCDPACK\footnote{http://www.astro.iag.usp.br/$\sim$antonio/gaveta/default.htm},
a collection of IRAF routines specifically developed to this purpose \citep{pereyra2000}.
This observational scheme provided mean polarization errors of approximately
$0.05$\%, so that typical interstellar polarization degree values ($0.1-2$\%)
could be easily obtained.

The observed stars were selected from the HIPPARCOS catalogue \citep{hipp97}, and
are distributed toward the interface region, mainly concentrated along
the proposed ring structure (Figure \ref{distr_data}, crosses). 
Our goal was to obtain a stellar distribution within
this feature with a maximum possible spatial uniformity. However, some observational limits
were imposed, e.g., notice a lack of observed data around $(l,b)\approx(300^{\circ},-30^{\circ})$, which
is very close to the south celestial pole's line-of-sight, and therefore could not be reached by
the telescope. Moreover, in order to observe the entire structure of the proposed ring,
our observational nights were roughly uniformly distributed along the year, allowing each
part of the feature to be observed during specific months. However, some regions coincided
with bad weather seasons at OPD, providing less observing time along these regions.
Besides observing along the ring, we have also chosen specific lines-of-sight in the direction
of the R CrA and Southern Coalsack clouds, which are located inside the contour of the ring.
A small fraction of the collected data are spread throughout these areas (see Figure \ref{distr_data}).

To guarantee
an appropriate distance precision up to $\sim 200$ pc, only stars with
trigonometric parallax relative error $\Delta \pi / \pi < 30$\% were used.
Additionally to our sample, we have also used the available polarimetric
data from \citet{heiles2000} catalogue, restricted to objects containing
distance data from the HIPPARCOS catalogue with $\Delta \pi / \pi < 30$\%.
Their distribution is also displayed in Figure \ref{distr_data}.
The whole sample, distributed along the region delimited by the Galactic coordinates
$240^{\circ} < l < 60^{\circ}$, $-60^{\circ} < b < 60^{\circ}$,
adds up to $2271$ stars, among which $1393$ are from \citet{heiles2000} catalogue
and $878$ objects were observed at OPD (using the Johnson's V filter).
The polarimetric data set is shown in Table \ref{poldata_short}.

   \begin{figure}[!h]
   \centering
   \includegraphics[width=0.5\textwidth]{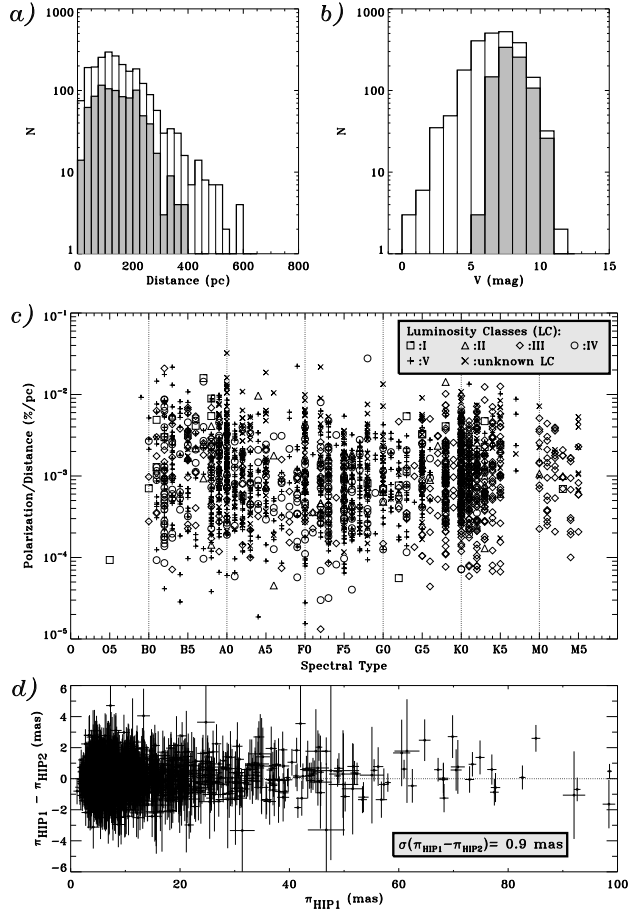}
      \caption{{\normalsize Histograms (a) and (b) respectively show the distributions of distance and
               Johnson's V magnitude, where filled columns represent only data
               collected at OPD and white columns consider the entire sample
               (including data from \citet{heiles2000}). Part (c) shows a diagram of 
               $\mathrm{P(\%)/d(pc)} \times \mathrm{Spectral\ Type}$ for the entire sample, 
               using logarithmic scale to improve visualization and distinguishing 
               the luminosity classes through different symbols, as shown at the gray box. 
               Most of the stars are distributed between B and K types, and no 
               $P/d$ trend can be seen related to giants or late-type stars, which could 
               be associated to intrinsic polarization.
               The diagram from part (d)
               represents a measure of trigonometric paralaxes differences $(\pi_{\mathrm{HIP1}}-\pi_{\mathrm{HIP2}})$
               between the Hipparcos original
               release \citep[HIP1,][]{hipp97} and the new reduction provided by \citet{van2007} (HIP2),
               using all stars from our sample. This comparison shows a small and {\it symmetric} dispersion
               around the equality line (the dotted line), with a standard deviation of $0.9$ milli-arcsecond (mas).}}
         \label{hist_data}
   \end{figure}

With respect to the distance distribution we have chosen stars that are mainly located
between $25$ and $250$ pc, as may be seen from the histogram shown
in part (a) of Figure \ref{hist_data}.
The filled portion of the histogram corresponds to the data collected at OPD, while the
white columns represent the entire sample (which includes the data from Heiles (2000)).
Based on the previous knowledge that the dense interstellar structures from the interface
would be located between $\sim80$ and $180$ pc, this represents a suitable distribution
allowing the detection of many of these clouds. Beyond $250$ pc there is a steep decrease in the
number of observed stars.

We have also plotted a histogram of the Johnson's V magnitude (Figure \ref{hist_data}b) 
for the entire sample (white portion) and more specifically
to the objects observed at OPD (filled parts). Notice that most of the stars observed
at OPD have $V > 6$, which is related to the saturation limit of the CCD used. Our sample
is limited to a maximum V magnitude of $\sim 11-12$ mag, i.e., most of the objects are bright
stars.

Figure \ref{hist_data}c shows a diagram of $\mathrm{P(\%)/d(pc)} \times \mathrm{Spectral\ Type}$,
where we have used a logarithmic scale in order to obtain a better visualization, 
and luminosity classes are indicated by different symbols.
Spectral types are mainly distributed between B and K types, reflecting the
general trend for objects of the Galactic disk. No trend for specific $P/d$ values is noted toward giants or
late type stars, which could indicate the presence of {\it intrinsic polarization}.
Concerning such topic, it is possible that some
of the objects of our sample present this property. However, we do not believe
that this would alter the general picture provided by our study.
Specifically, as will be shown in sections \ref{polvec} and \ref{poldist},
it is improbable that intrinsic polarization could explain the large-scale correlation of polarization angles,
and also the correlated distances where a rise in polarization degree and color excess is seen
in the direction of several regions. 
Using the SIMBAD\footnote[1]{http://simbad.u-strasbg.fr/simbad/} object classification tool
for the entire sample, only six stars are designated pre-main sequence stars, namely:
HIP 80337, HIP 80462, HIP 6494, HIP 93368, HIP 64408 and HIP 110778.
Among these, HIP 6494 and HIP 93368 were observed at OPD, while the others were obtained
from \citet{heiles2000}.

   \begin{figure}[!b]
   \centering
   \includegraphics[width=0.5\textwidth]{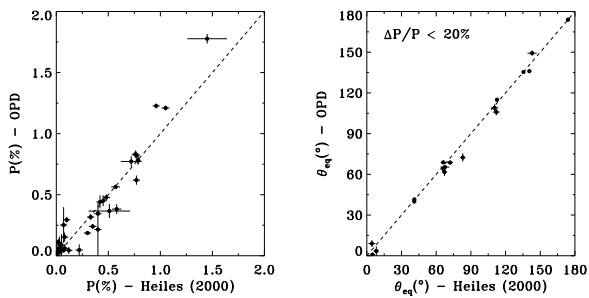}
      \caption{Comparison between polarimetric data from \citet{heiles2000} catalogue and
               those obtained at OPD for some objects of the sample. The
               dashed lines were inserted for clarity, and for the $\theta$ (equatorial) comparison,
               only objects with $\Delta P/P < 20$\% were included in order to exclude the
               unpolarized stars.
              }
         \label{heiles_opd_comparison}
   \end{figure}

We have used the original data from Hipparcos (HIP1, \citet{hipp97}), instead of the new reduction data provided
by \citet{van2007} (HIP2).
Since a comparison of trigonometric parallaxes between both catalogs reveal a small
and symmetric dispersion around the equality line for the stars of our sample (see Figure \ref{hist_data}d),
the analysis is not affected by the new reduction scheme, i.e., 
if small shifts would increase and decrease the distances of
individual stars, this would not change the general picture provided by our results.

As a means of verifying the consistency of the collected data, some of the
observed stars were chosen to match objects
available from \citet{heiles2000} catalogue. Figure \ref{heiles_opd_comparison}
shows a comparison of the polarization degree and angle for both samples.
The mean percentage difference between the compared values is only $\sim13$\% for
$P_{V}$ and $\sim16$\% for $\theta$.

\section{Correlation between $P(\%)$ and $E(b-y)$}
\label{correlation_p_colourexcess}

\hspace{1.0em} Whenever polarization degree is plotted against interstellar reddening,
it is observed to exist a maximum value of $P$, empirically corresponding
to the inequality  $P_{V}/E(B-V) <  9.0 \% \mathrm{mag}^{-1}$
\citep{serkowski_etal_1975}.
This comes from the fact that (after the passage through a successive number
of interstellar clouds) reddening of stellar light is an additive effect,
whereas polarization exhibits a more complex behavior, heavily dependent on the
alignment efficiency.

   \begin{figure}[!b]
   \centering
   \includegraphics[width=0.5\textwidth]{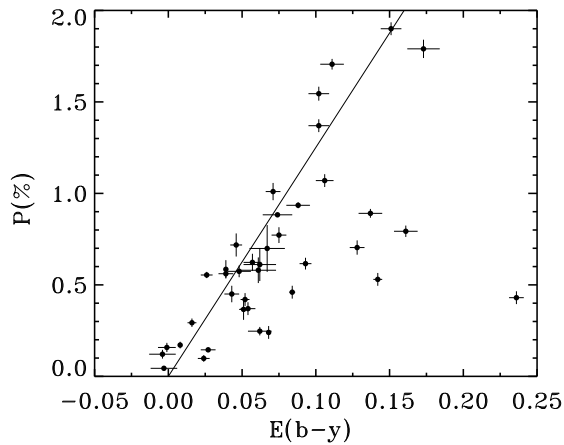}
      \caption{Diagram showing the limited correlation between
               $P(\%)$ and $E(b-y)$. The straight line corresponds to the ideal
               conditions of alignment efficiency ($P_{V}/E(b-y) = 12.51$) and the photometric data were
               obtained from \citet{reis_corradi2008}.
              }
         \label{relation_P_excess}
   \end{figure}

It is then instructive to verify this relation to the objects of our data, using stars
to which color excesses have been accurately determined. This could be achieved
by correlating our sample to the $E(b-y)$ data of \citet{reis_corradi2008}, who
studied the same interaction region using Str\" omgren photometry. The results
are shown in Figure \ref{relation_P_excess}, where the straight line corresponds
to the optimum alignment efficiency, which is reached only under ideal conditions.
This relation, expressed by $P_{V}/E(b-y) = 12.51$, was obtained by using
$E(B-V)=1.39E(b-y)$ with the empirical relation mentioned above.
Although several stars show a ratio $P_{V}/E(b-y) < 12.51$, as expected, we can see that
most of them lie close to the optimum alignment line, indicating a quite good
correlation between these two quantities.

It is possible to find a star that shows a high value of color excess,
and a low polarization degree, which in our case would represent a problem, as we
expect to detect a rise in polarization wherever an interstellar structure is localized.
However, in order to affect our analysis, a systematic decrease in alignment efficiency
would have to be observed over large areas of the sky ($>10$ square degrees).
Therefore, the detection of a polarized star implies a minimum expected value of color
excess and hence, when studied as a function of distance, a rise in polarization
in the direction of an specific interstellar structure may be interpreted as
setting a superior limit to the distance of that cloud.

\section{Distribution of the polarization vectors toward the interface region}
\label{polvec}

   \begin{figure*}[!t]
   \includegraphics[width=\textwidth]{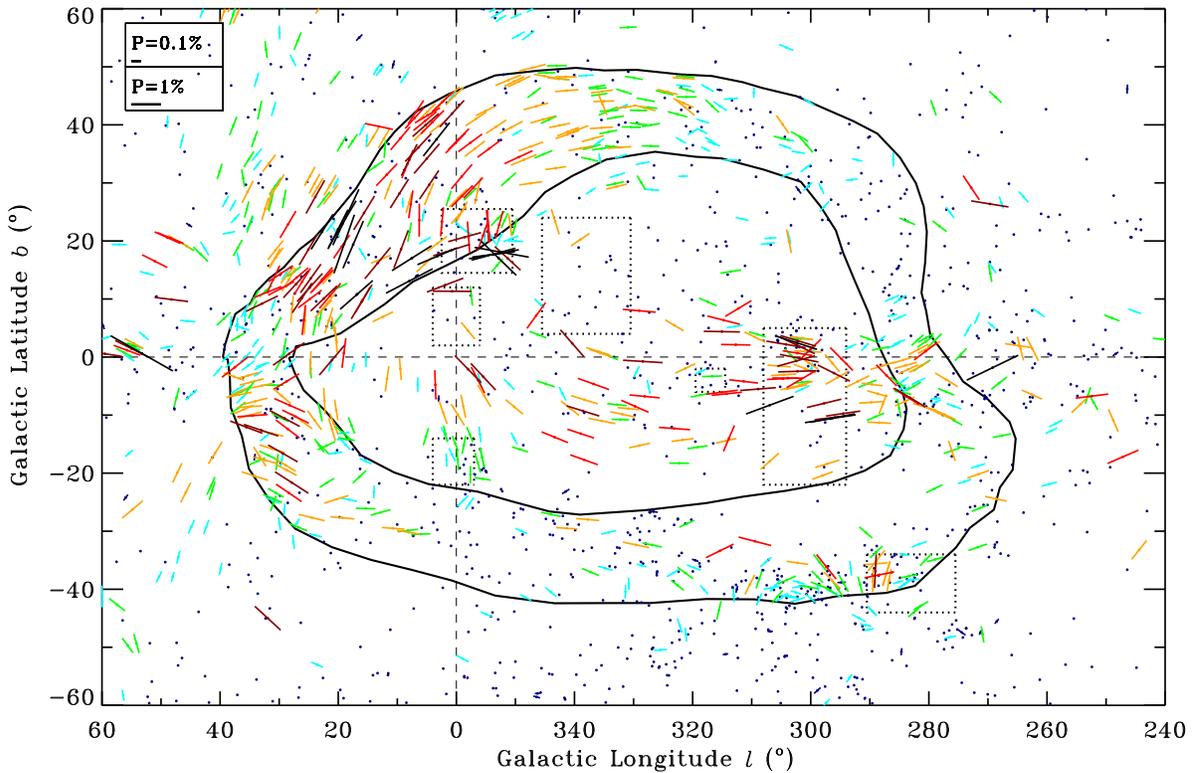}
      \caption{Distribution of polarization vectors along the studied region. The thick
               contour corresponds to the annular feature proposed by \citet{egger_aschenbach1995}
               and the dotted rectangles indicate the locations of the main dark clouds distributed along
               this region, as shown in Figure \ref{distr_data}. To improve visualization,
               the sizes of the $P$-vectors are plotted proportionally to $\sqrt{P}$, and a grayscale scheme
               (colored at the electronic version) was used
               (as specified on Table \ref{color_pol}).
              }
         \label{vecpol_integrated}
   \end{figure*}

\hspace{1.0em}
The model that suggests the existence of a dense and neutral ``wall" of interstellar material between
the LC and Loop I claims that the supposed interaction ring would be located at a distance
of approximately $70$ pc, corresponding to a $N_{\mathrm{HI}}$ transition from
$\sim 10^{20}$ cm$^{-2}$ to $\sim 7\times 10^{20}\ $cm$^{-2}$ \citep{egger_aschenbach1995}.
In this case, the standard gas-to-dust ratio proposed by \citet{knude1978} may be used
to interpret this transition as a rise in $E(b-y)$ from $\approx0\fm015$ to $\approx0\fm100$, which
corresponds to maximum value of polarization ($P_{max}$) increasing from $\approx0.2$ to $\approx1.3\%$.

\begin{table}[!h]
\caption
{Color/Grayscale scheme used in the polarization analysis}
\begin{tabular}{cccc}
\hline\hline
$P(\%)$ & Color & Grayscale & $E(b-y)_{min}$ \\ \hline
$\leq 0.1$ & {\color{NavyBlue} navy blue} &  black & $\leq0\fm008$ \\
$0.1-0.3$  & {\color{Cyan} cyan} & light gray &  $\approx0\fm008-0\fm025$ \\
$0.3-0.5$  & {\color{Green} green} & light gray &  $\approx0\fm025-0\fm040$ \\
$0.5-0.9$  & {\color{Orange} orange} & light gray &  $\approx0\fm040-0\fm070$ \\ \hline
$\mathbf{0.9-1.3}$  & {\bfseries{\color{Red} red}} & dark gray &  $\mathbf{\approx0\fm070-0\fm100}$ \\ \hline
$1.3-2.0$  & {\color{BrickRed} dark red} & black &  $\approx0\fm100-0\fm160$ \\
$> 2.0$    & {\color{Black} black} & black &  $> 0\fm160$ \\  \hline
\end{tabular}
\tablecomments{The colors (shown at the electronic version) and grayscale levels above are related
to each polarization interval, as well as the corresponding
minimum color excess value ($E(b-y)_{min}$), assuming that the $P/E(b-y)<12.51$ relation is valid.
The dark gray level (i.e., the red color) represent the expected transition along the annular region.
}
\label{color_pol}
\end{table}

   \begin{figure}[!h]
   \centering
   \includegraphics[width=\textwidth]{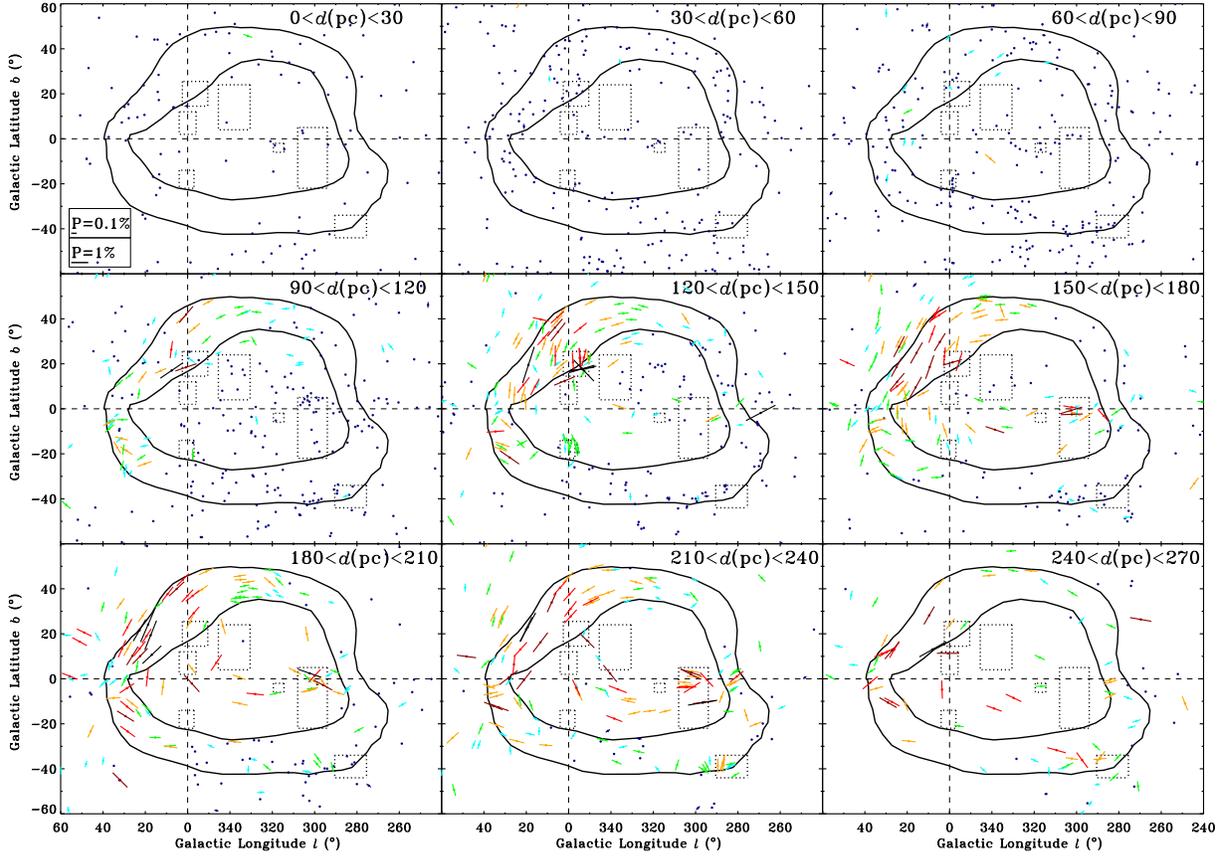}
      \caption{Distribution of polarization vectors along the interface region, divided
               by distance intervals of $30$ pc each. Notice the existence of dark gray vectors
               (colored in red at the electronic version, corresponding to the
               transition to $P\approx0.9-1.3\%$) on the left side at about $90 - 120$pc (the loop-shaped polarization
               feature in the direction of $(l,b)\approx(0^{\circ},30^{\circ})$), whereas on the right side
               they are not clearly seen before $240$pc. The details are the same as in Figure \ref{vecpol_integrated}.
             }
         \label{vecpol_intervals}
   \end{figure}

\citet{knude1978} asserts that stars with $E(b-y)\geq0\fm030$ must be located behind at least
a diffuse interstellar cloud. Hence, if $E(b-y)\sim0\fm070$,
the objects are probably reddened by a denser interstellar cloud, which might as well polarize the stellar light.
Based on this reasoning, in the following polarization analysis we shall look for a transition
from $P_{max}\approx0.2\%$ to $P_{max}\approx0.9-1.3\%$, which corresponds to $E(b-y)$
going from $0\fm015$ to $0\fm070-0\fm100$.
In principle, this is the expected transition if the proposed HI column density for the ring is considered.
To ease the comparison between the polarization and the color excess we will adopt a
criterion similar to \citet{reis_corradi2008}, for the typical values of $E(b-y)$ corresponding
to the ring transition.

The distribution of polarization vectors along the entire studied region is shown on Figure
\ref{vecpol_integrated}, where lines represent the $\mathbf{\bf{E}}$-vectors of the partially
polarized light, centered on the position of each star. We have chosen to represent
the size of each vector proportional to $\sqrt{P}$, hence allowing a better visualization
of the polarization pattern, since most of the objects show low polarization degree ($\leq 0.5\%$).
For polarized stars ($P\geq0.1\%$), only those satisfying $\Delta P/P < 0.2$ and $\Delta\theta_{gal} < 20^{\circ}$
are plotted, eliminating 338 objects from this kind of diagram. 
The $\Delta\theta_{gal} < 20^{\circ}$ requirement implies that objects
with large uncertainties on polarization angle will not be plotted.
Considering the typical polarization degree error of $\sim 0.05\%$, we define objects with
$P\leq0.1\%$ as ``unpolarized", based on our instrumental precision. These objects
are represented as a single dot.

Moreover, in order to further improve the visualization,
a color scheme corresponding to polarization degree intervals was adopted.
Table \ref{color_pol} specifies the colors and grayscales corresponding to each polarization interval,
as well as the minimum color excess value ($E(b-y)_{min}$) associated to each $P$
(assuming that the relation discussed on section \ref{correlation_p_colourexcess} is valid).
Notice that the dark gray level (i.e., the red color at the electronic version) corresponds to the transition of $0.9-1.3\%$.

\placetable{color_pol}

To analyze the distribution of the polarization vectors as a function of
distance, the same diagram of Figure \ref{vecpol_integrated} was sliced in distance intervals of $30$ pc each.
These plots are shown on Figure \ref{vecpol_intervals}, from $0$ to $270$ pc.
Regions that are closer to the Sun (up to $60$ pc) exhibit mainly unpolarized stars ($P<0.1\%$),
consistent with the presence of the LC. The first weakly polarized stars (with $P<0.5$\%)
are distributed along the western (left) side ($l>330^{\circ}$) of the interaction region in the $60 - 90$ pc interval.
The first large-scale interstellar structure is observed in the $90 - 120$ pc interval near
$(l,b)\approx(0^{\circ},30^{\circ})$,
where the polarization vectors run parallel to the ring contour and show high polarization
degree ($P>1.3\%$) along most lines-of-sight, forming something like a ``loop" pattern.
Notice that up to this distance range, the eastern (right) side ($l\approx 290^{\circ}$) is still composed
mainly by unpolarized stars.

The main dark clouds located inside the ring contour are located along the $120 - 150$ pc and $150 - 180$ pc
distance intervals, e.g.: R CrA ($l,b\approx0^{\circ},-18^{\circ}$), $\rho$ Oph ($l,b\approx355^{\circ},+20^{\circ}$)
and the Southern Coalsack/Chamaeleon/Musca complex ($l,b\approx300^{\circ},-10^{\circ}$).
Notice that the large-scale structure observed along the left side
of the ring shows polarization vectors distributed roughly
parallel to the ring contour orientation. The first few weakly polarized stars ($P<0.3$\%) appear at
the south-right side ($280^{\circ}<l<320^{\circ}$,$-50^{\circ}<b<-20^{\circ}$) in the $150 - 180$ pc interval.
However, the majority of the objects in this region are still unpolarized.

The $180 - 210$ pc and $210 - 240$ pc diagrams show that the first stars
located along the right side of the ring presenting slightly higher polarization degree 
($P>0.5$\%) appear only after $200$pc.
Also notice that the polarization vectors along this direction are
perpendicular to the ring structure, in contrast to the situation observed at the
left side, where the vectors run parallel to the ring contour.

The $240 - 270$ pc diagram shows that up to $270$ pc the right region
is composed predominantly by stars with polarization in the range $0-0.9$\%.
Some stars possess $P>0.9\%$ suggesting that the expected transition to
$P\sim0.9-1.3\%$ might occur along this distance range.
For higher distances, the sample becomes non-uniform and incomplete,
an expected effect, given the rapid decrease in the number
of observed stars beyond $250$pc, as can be seen from Figure \ref{hist_data}a.

\section{Polarization as a function of distance at the interface region}
\label{poldist}

\hspace{1.0em}
Several interstellar structures are distributed along the interface with the Loop I superbubble,
and can be identified by a rise in the $P(\%)$ values when such a feature is crossed.
To build a better picture of the distances to each individual structure
along the whole region, we have plotted diagrams of $P(\%)$ as a function of distance $d$(pc)
to several rectangular areas positioned along the entire surveyed area. These diagrams are limited
to the range $0\%<P<2\%$ to offer a better view of the smaller $P$ variations.

   \begin{figure}[!t]
   \centering
   \includegraphics[width=\textwidth]{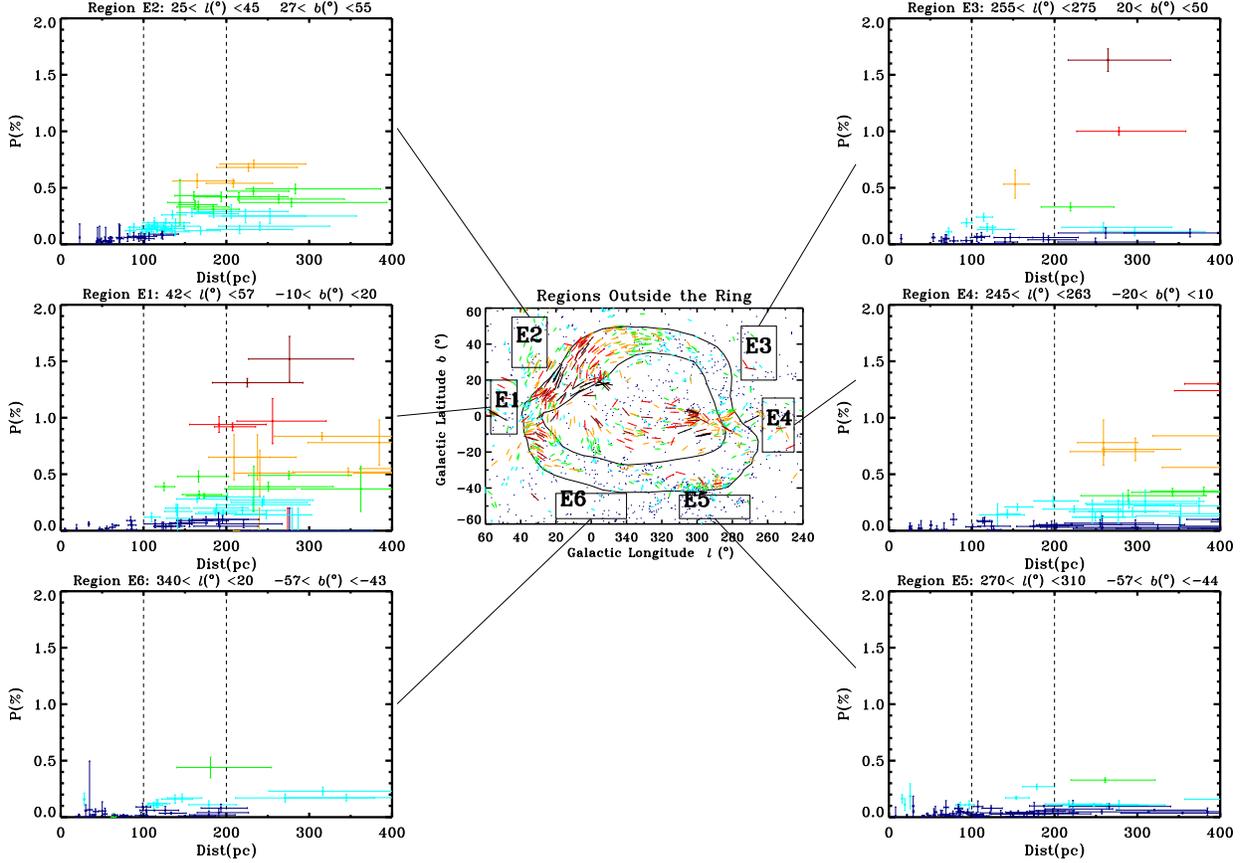}
      \caption{Diagrams of $P(\%) \times distance(pc)$ for rectangular
               regions outside the ring, as indicated in the central panel.
               Polarization vectors are also plotted along the interface
               region at the center, as in Figure \ref{vecpol_integrated}.
               A rise to $P\approx 0.6$\% is observed at approximately $130-140$ pc along E1 and E2 areas,
               coinciding to the direction of the NPS.
               These diagrams are limited to the range $0\%<P<2\%$.
              }
         \label{pdist_exterior}
   \end{figure}

\begin{figure}[!t]
   \centering
   \includegraphics[width=0.9\textwidth]{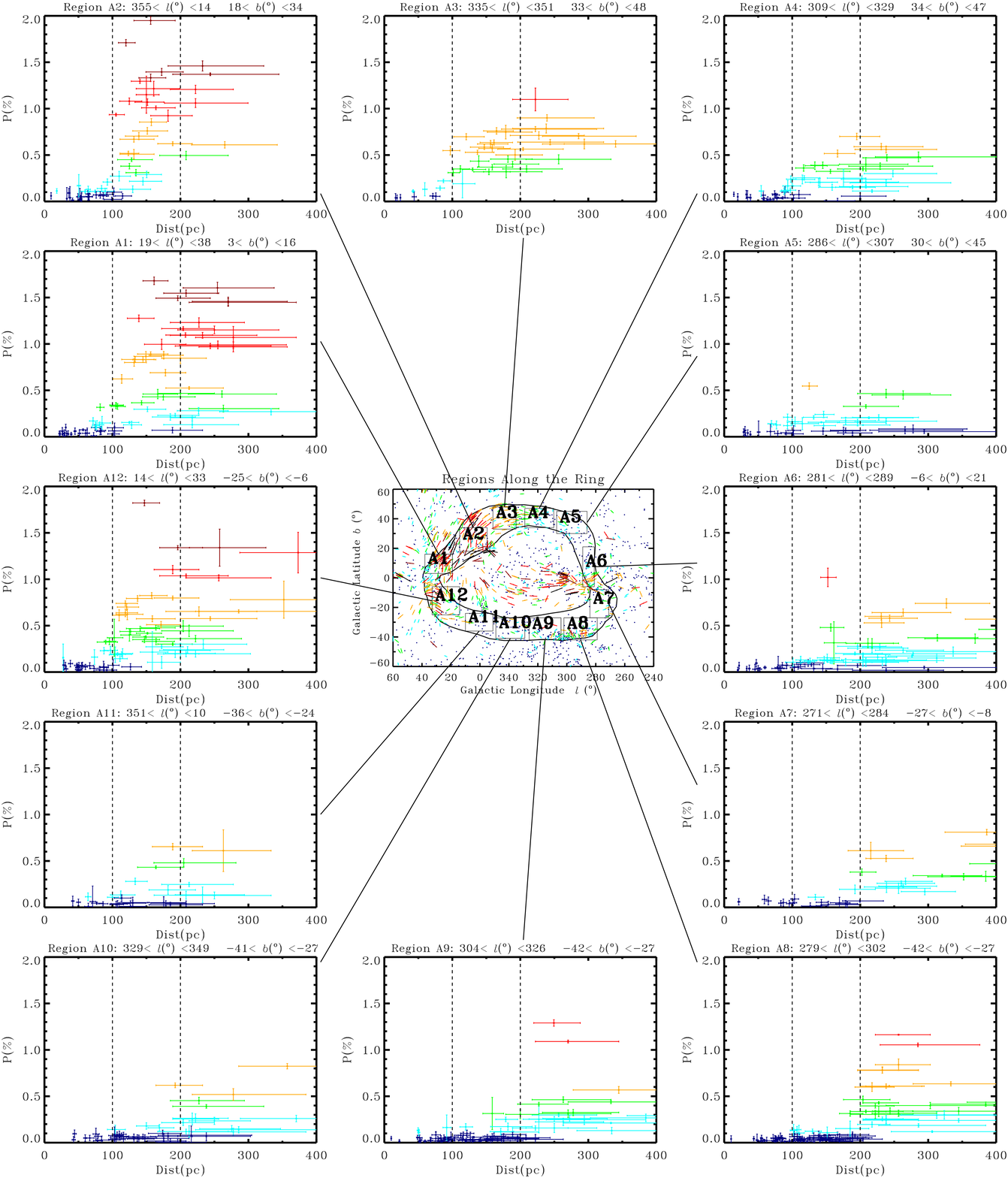}
      \caption{Diagrams of $P(\%) \times distance(pc)$ for rectangular
               regions along the ring. The $0.9-1.3\%$ transition occurs at $\approx100$ pc
               on the left side (e.g., A2 area) and is not clearly identified before
               $\approx250$ pc on the right side region (e.g., A8 area).
               These diagrams are limited to the range $0\%<P<2\%$.
              }
         \label{pdist_ring}
   \end{figure}

\begin{figure}[!t]
   \centering
   \includegraphics[width=\textwidth]{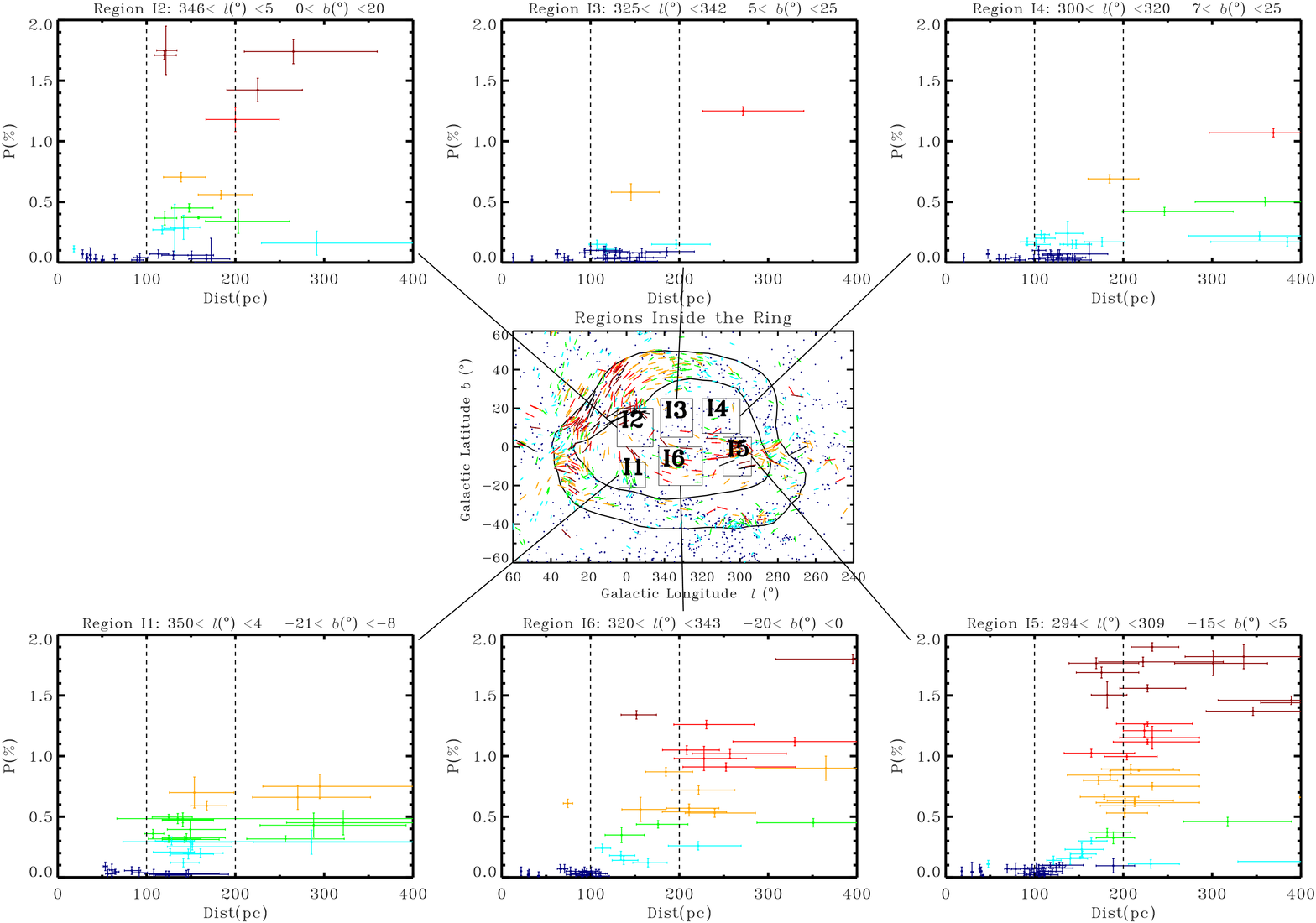}
      \caption{Diagrams of $P(\%) \times distance(pc)$ for rectangular
               regions inside the ring.
               A rise in P(\%) is seen at $\approx 120-130$ pc in the directions of
               R CrA (region I1, $P_{max}\approx0.6\%$) and $\rho$ Oph (region I2, $P_{max}>2\%$).
               The complex of dark clouds defined by the Southern Coalsack, Chamaeleon, and Musca
               (I5 region) displays a sharp rise to $P>2\%$ at approximately $150-180$ pc.
               These diagrams are limited to the range $0\%<P<2\%$.
              }
         \label{pdist_interior}
   \end{figure}

Our analysis was divided between three regions: outside (E1-E6, Figure \ref{pdist_exterior}),
along (A1-A12, Figure \ref{pdist_ring}), and inside (I1-I6, Figure \ref{pdist_interior}) the annular
ring-like feature proposed by \citet{egger_aschenbach1995}. This scheme shall allow
us to study specifically the structure of the supposed ring and its relation to the
surroundings, as well as to compare our results with the previous photometric analysis
by \citet{reis_corradi2008}. The limiting Galactic coordinates of each rectangular region
are indicated in the corresponding figures.

\subsection{Regions {\it outside} the ring}

%

\hspace{1.0em}
For the regions outside the ring area (Figure \ref{pdist_exterior}),
interstellar polarization is mainly $\leq 0.3$\%
up to approximately $200$pc in almost all directions,
although more data would be necessary to carry out an adequate analysis.

Exceptions are observed at E1 and E2 regions, where a rise
to $P\approx 0.6$\% is observed at approximately $130-140$ pc.
These regions coincide with the direction of the ``North Polar Spur'' (NPS),
the brightest rim of Loop I, well-known to be a large radio filament,
which extends to the north along $30^{\circ}$ in longitude.
The rise in $P(\%)$ is consistent with the presence of this structure.

\subsection{Regions {\it along} the ring}

\hspace{1em}
A sharp rise in $P$ (in some cases higher than $1.3\%$) is observed for several areas along
the left side of the interaction region (areas A1, A2, A3, and A12 of Figure \ref{pdist_ring})
indicating that the interstellar structures are positioned at a distance of $\approx80-100$pc in these directions.
These areas coincide with the large loop pattern identified by analyzing the polarization vectors,
as discussed on Section \ref{polvec} (see, for example, Figures \ref{vecpol_integrated} and \ref{vecpol_intervals}).
Some dark clouds and dense structures are known to be present in these directions, particularly:
Sag-South and Aql-South ($27^{\circ}<l<40^{\circ}$; $-21^{\circ}<b<-10^{\circ}$), the Scutum dark cloud
($l=25^{\circ}$;$b=1^{\circ}$), the Oph-Sgr molecular clouds ($8^{\circ}<l<40^{\circ}$; $9^{\circ}<b<24^{\circ}$) etc.

Carrying on the analysis in the left-right direction,
a smooth decrease to $P\approx0.5\%$ is observed at the northern (e.g., A4 and A5)
and southern areas (e.g., A10 and A11) at $\sim100$ pc,
when compared to polarization values at the same distance on the left areas
(e.g., A1, A2, and A12).

In the regions to the right, the first clear rise to $P\approx0.6\%$ occurs at $180-200$ pc along
the A6, A7 and A8 areas. However, the expected $0.9-1.3\%$ transition is not clearly seen before $\sim250$ pc
(see A8 and A9 areas). Specifically, part of the A8 region coincides with a diffuse interstellar
filament in the direction of the Mensa constellation, with an estimated distance of
$d=230\pm30$ pc \citep{penprase1998}.

Some regions (e.g., the A5 region) show low polarization values ($P<0.5\%$)
up to $\sim300$ pc, suggesting that the structure is discontinuous and probably highly fragmented.

\subsection{Regions {\it inside} the ring}

\hspace{1em}
The regions inside the ring area (Figure \ref{pdist_interior}) are marked by the presence of
several well-known dark clouds distributed relatively close
to the GP along the southern sky \citep[see e.g.,][]{dame2001}.

The diagrams related to left areas exhibit a rise in P(\%) at $\approx 120-130$ pc in the directions of
R CrA (region I1, $P_{max}\approx0.6\%$) and $\rho$ Oph (region I2, $P_{max}>2\%$).
The I2 area also encompasses the region of the Pipe Nebula, which is located
at a distance of $145\pm16$ pc from the Sun \citep{alves2007}.

Along the I3 and I4 regions our data are quite sparse and do not lead us to
any clear conclusion, although there seems to be a trend to low polarization
values ($P<0.6\%$) up to $\approx200$ pc.
The complex of dark clouds defined by the Southern Coalsack, Chamaeleon, and Musca
(SCCM, I5 region) displays a sharp rise to $P>2\%$ at approximately $150-180$ pc.
This distance corroborates previous photometric analysis by \citet{corradi1997,corradi2004}.
A smooth rise from $P<0.1\%$ (up to $\sim100$ pc) to $P\approx1.3\%$ (at $\sim200$ pc)
is observed at the I6 area.

\section{Discussion}
\label{discussion}

\subsection{Summary of polarimetric properties between regions outside, along and inside the ring}

\begin{table}[!t]
\scriptsize
\caption
{Summary of the analysis of rectangular areas outside, along, and inside the ring}
\begin{tabular}{cccc}
\hline\hline
Area & $(l_{min},l_{max})$ & $(b_{min},b_{max})$ & Comment \\ \hline
E1  & $(42 ^{\circ},57 ^{\circ})$ & $(-10^{\circ},20 ^{\circ})$ & Orientation along the NPS; $P$ rise to $\approx 0.6$\% at $130-140$ pc \\
E2  & $(25 ^{\circ},45 ^{\circ})$ & $(27 ^{\circ},55 ^{\circ})$ & Same as E1; Northern portion of the NPS loop \\
E3  & $(255^{\circ},275^{\circ})$ & $(20 ^{\circ},50 ^{\circ})$ & Low $P$ ($\leq 0.3$\%) up to approximately $200$ pc \\
E4  & $(245^{\circ},263^{\circ})$ & $(-20^{\circ},10 ^{\circ})$ & Same as E3 \\
E5  & $(270^{\circ},310^{\circ})$ & $(-57^{\circ},-44^{\circ})$ & Same as E3 \\
E6  & $(340^{\circ},20 ^{\circ})$ & $(-57^{\circ},-43^{\circ})$ & Same as E3 \\ \hline
A1  & $(19 ^{\circ},38 ^{\circ})$ & $(3   ^{\circ},16 ^{\circ})$ & $P$ rise to $\approx 0.7$\% at $80-100$ pc; loop pattern; Serpens-Aquila cloud \\
A2  & $(355^{\circ},14 ^{\circ})$ & $(18  ^{\circ},34 ^{\circ})$ & $P$ rise to $>1.0$\% at $\approx 100$ pc; loop pattern orientation \\
A3  & $(335^{\circ},351^{\circ})$ & $(33  ^{\circ},48 ^{\circ})$ & $P$ rise to $\approx 0.7$\% at $\approx 100$ pc; $P$ decrease compared to A1 and A2 \\
A4  & $(309^{\circ},329^{\circ})$ & $(34  ^{\circ},47 ^{\circ})$ & Same as A3; final portion of the polarization loop pattern \\
A5  & $(286^{\circ},307^{\circ})$ & $(30  ^{\circ},45 ^{\circ})$ & Low $P$ values ($<0.5\%$) up to $\sim300$ pc \\
A6  & $(281^{\circ},289^{\circ})$ & $(-6  ^{\circ},21 ^{\circ})$ & Orientation parallel to the GP; $P$ rise to $0.6$\% at $\approx 180-200$ pc \\
A7  & $(271^{\circ},284^{\circ})$ & $(-27 ^{\circ},-8 ^{\circ})$ & $P$ rise to $0.6$\% at $\approx 180-200$ pc \\
A8  & $(279^{\circ},302^{\circ})$ & $(-42 ^{\circ},-27^{\circ})$ & $P=0.9-1.3\%$ at $\sim250$ pc; interstellar filament at $d=230\pm30$ pc \\
A9  & $(304^{\circ},326^{\circ})$ & $(-42 ^{\circ},-27^{\circ})$ & $P=0.9-1.3\%$ transition at $\sim250$ pc \\
A10 & $(329^{\circ},349^{\circ})$ & $(-41 ^{\circ},-27^{\circ})$ & $P$ rise to $0.6$\% at $\approx 180-200$ pc \\
A11 & $(351^{\circ},10 ^{\circ})$ & $(-36 ^{\circ},-24^{\circ})$ & $P$ rise to $0.6$\% at $\approx 150$ pc; $P$ decrease compared to A12 \\
A12 & $(14 ^{\circ},33 ^{\circ})$ & $(-25 ^{\circ},-6 ^{\circ})$ & $P$ rise to $\approx 0.7$\% at $80-100$ pc; \\ \hline
I1  & $(350^{\circ},4  ^{\circ})$ & $(-21^{\circ},-8^{\circ})$ & R CrA dark cloud; rise to $P\approx0.6\%$ at $\approx 120-130$ \\
I2  & $(346^{\circ},5  ^{\circ})$ & $(0  ^{\circ},20^{\circ})$ & $\rho$ Oph and Pipe Nebula dark clouds; rise to $P>2\%$ at $\approx 120-130$ \\
I3  & $(325^{\circ},342^{\circ})$ & $(5  ^{\circ},25^{\circ})$ & Low $P$ ($<0.6\%$) up to $\approx200$ pc; few data \\
I4  & $(300^{\circ},320^{\circ})$ & $(7  ^{\circ},25^{\circ})$ & Same as I3 \\
I5  & $(294^{\circ},309^{\circ})$ & $(-15^{\circ},5 ^{\circ})$ & SCCM dark clouds; rise to $P>2\%$ at $\approx 150-180$ pc \\
I6  & $(320^{\circ},343^{\circ})$ & $(-20^{\circ},0 ^{\circ})$ & Smooth rise from $P<0.1\%$ (at $\sim100$ pc) to $P\approx1.3\%$ (at $\sim200$ pc) \\ \hline
\end{tabular}
\tablecomments{
The first column indicates the name of each rectangular area associated to figures \ref{pdist_exterior}, \ref{pdist_ring}, and \ref{pdist_interior},
divided by regions outside, along and inside the ring contour (labeled ``E", ``A", and ``I", respectively).
The second and third columns respectively indicate the limits (minimum and maximum) of $l$ and $b$ coordinates associated
to each area. The fourth column shows a short summary of the main polarimetric property along
each region.}
\label{regions_summary}
\end{table}

The main polarimetric properties exposed in sections \ref{polvec} and \ref{poldist}, 
associated to each of the rectangular areas inside, along and outside the ring contour
are summarized in table \ref{regions_summary}.

Inside the contour, large polarization values ($P>2\%$) are seen in the direction of several 
dark clouds distributed along the interface area. When analyzing the distances to these interstellar 
structures, we notice a difference in positions of clouds from the left side (R CrA, $\rho$ Oph; at 
$120-130$ pc) when compared to the right side (Southern Coalsack, Chamaeleon and Musca complex; at
$150-180$ pc). This may indicate that the interface is tilted with respect to the Sun. 

Figure \ref{vecpol_integrated} show that inside the contour the polarized vectors' orientation is mainly 
parallel to the GP. However, along the left side (particularly along the R CrA dark cloud), vectors 
are oriented perpendicularly to the GP. According to \citet{harju1993}, an expanding shell
from the Upper Centaurus-Lupus (UCL) Sco-Cen sub-group would have collided with the R CrA dark cloud
a few million years ago, creating the cloud's cometary shape and triggering star formation. 
In this direction, polarized vectors are oriented parallel to the ``head-tail" 
direction of the cloud, and therefore perpendicular to the GP. 
This could be a local effect of the distortion of magnetic field lines due to the 
passage of the expanding shell.

Along the contour of the ring, a coherent alignment of polarization vectors appears toward 
the left-northern side (areas A1, A2, A3 and A4), tracing a loop-shaped structure of magnetic field lines.
The distance to this structure is evident from a rise in polarization values 
at $80-100$ pc. 
The Serpens-Aquila molecular cloud ($l\approx30^{\circ},b\approx4^{\circ})$, which is positioned toward the A1 area, 
is probably located at $200$ pc, according to \citet{dame1987,dame2001}. In fact, the A1 diagram shows a  
second polarization rise at $\sim 200$ pc, which could correspond to this interstellar structure. However, the first rise to 
$P\approx0.6\%$ at $\sim100$ pc (also detected at the photometric survey) suggests the 
presence of nearer interstellar material in front of the denser molecular cloud.

Along the right side of the ring, the first rise to low polarization values
$\sim 0.5\%$ occurs only after $200$ pc. Furthermore, the vectors' orientation is mainly
uncorrelated to the contour of the ring.
Further discussion of vectors' orientation along the ring will be presented in section \ref{disc_polvec_ring}

Outside the contour of the ring the most conspicuous structure is the North Polar Spur (along E1 and E2 areas),
located at approximately $130-140$ pc. In this direction polarized vectors are 
oriented in a loop pattern parallel to the radio filament.

\subsection{Comparison with the color excess analysis}
\label{comp_ex}

\hspace{1em}
The results presented in section \ref{poldist} are in agreement with a similar analysis
made by \citet{reis_corradi2008} using $E(b-y)$ color excesses as a function of distance to several rectangular
areas located along the interface region. Particularly, the above mentioned work
revealed that the expected transition to $E(b-y)\approx0\fm070-0\fm100$, corresponding
to the ring's column density, occurs on the left side of the ring
at $d=110\pm20$ pc, while the right side transition
is not clearly seen before $d=280\pm50$ pc. Accordingly, our data show that the left side rise to
$P\sim0.9-1.3\%$ occurs at $\approx100$ pc, while the right side regions exhibit the same
transition only beyond $\approx250$ pc. 

The results of both surveys also match for areas inside and outside the ring region.
When comparing the polarimetric and photometric analysis along each line-of-sight, the increase in polarization
and $E(b-y)$ always occur at the same distances. For example, if we compare Figures 9, 10 and 11 from
\citet{reis_corradi2008} with our Figures \ref{pdist_exterior}, \ref{pdist_ring} and \ref{pdist_interior},
we note that: in the direction of A2, a rise
both in polarization and $E(b-y)$ is seen at approximately $100$ pc;
toward the SCCM clouds (I5 area), both $P$ and $E(b-y)$ increases sharply at $150-180$ pc.
Such correlations may be noted throughout several other areas and may be explained by the good association between
color excesses and polarization values, as presented on section \ref{correlation_p_colourexcess}.

Despite the correlation between $P$ and $E(b-y)$, polarization efficiency variations along the 
ring possibly could exist locally,
toward specific areas, particularly in the direction of star-forming regions, where the radiation field
is more intense \citep{andersson2007}.
However, given the largely correlated distances obtained from the polarimetric analysis when compared
to the photometric survey \citep{reis_corradi2008}, {\it local} variations of $P/A_{V}$ does
not affect our {\it global} conclusions on the positions of the several interstellar structures.

Comparing the polarization properties between the interfaces's left and right sides,
it is worth noting that the gradual decrease in the highest polarization values observed in the
right side ($P\approx2\%$) relative to the left ones ($P\approx0.9\%$) may be due to
a greater proximity of the right side with the ``magnetic pole" associated to the large-scale
structure of the Galactic magnetic field at the solar neighborhood, which is mainly
directed along the local spiral arm of the Galaxy. According to \cite{whittet2003} and \cite{heiles2005}
the magnetic pole is located toward $(l,b)=(260^{\circ},0^{\circ})$, and can be viewed face-on in this direction.
This is revealed by a greater dispersion
in polarization angles and lower polarization degree values toward this line-of-sight, which traces only
the sky-projected component of the magnetic field. However, this fact doesn't impair
our analysis of polarization as a function of distance toward the right side regions,
since the same results are observed with the color excess analysis (which does not
depend on the magnetic field configuration), revealing similar distances to the
interstellar structures.

These evidence confirm the distorted nature of the
large-scale interface of the interstellar medium in the direction of Loop I. Besides,
by analyzing specifically the A5 and A11 areas along the ring, we note a lower dust
column density along these lines-of-sight (see Figures \ref{vecpol_integrated} and \ref{pdist_ring})
which may be an indication of fragmentation of the suggested large-scale ring structure,
or simply that such ring feature may not exist.

\subsection{Analysis of the polarization vectors' direction and its relation to the ring structure}
\label{disc_polvec_ring}

\begin{figure}[!t]
   \centering
   \includegraphics[width=0.8\textwidth]{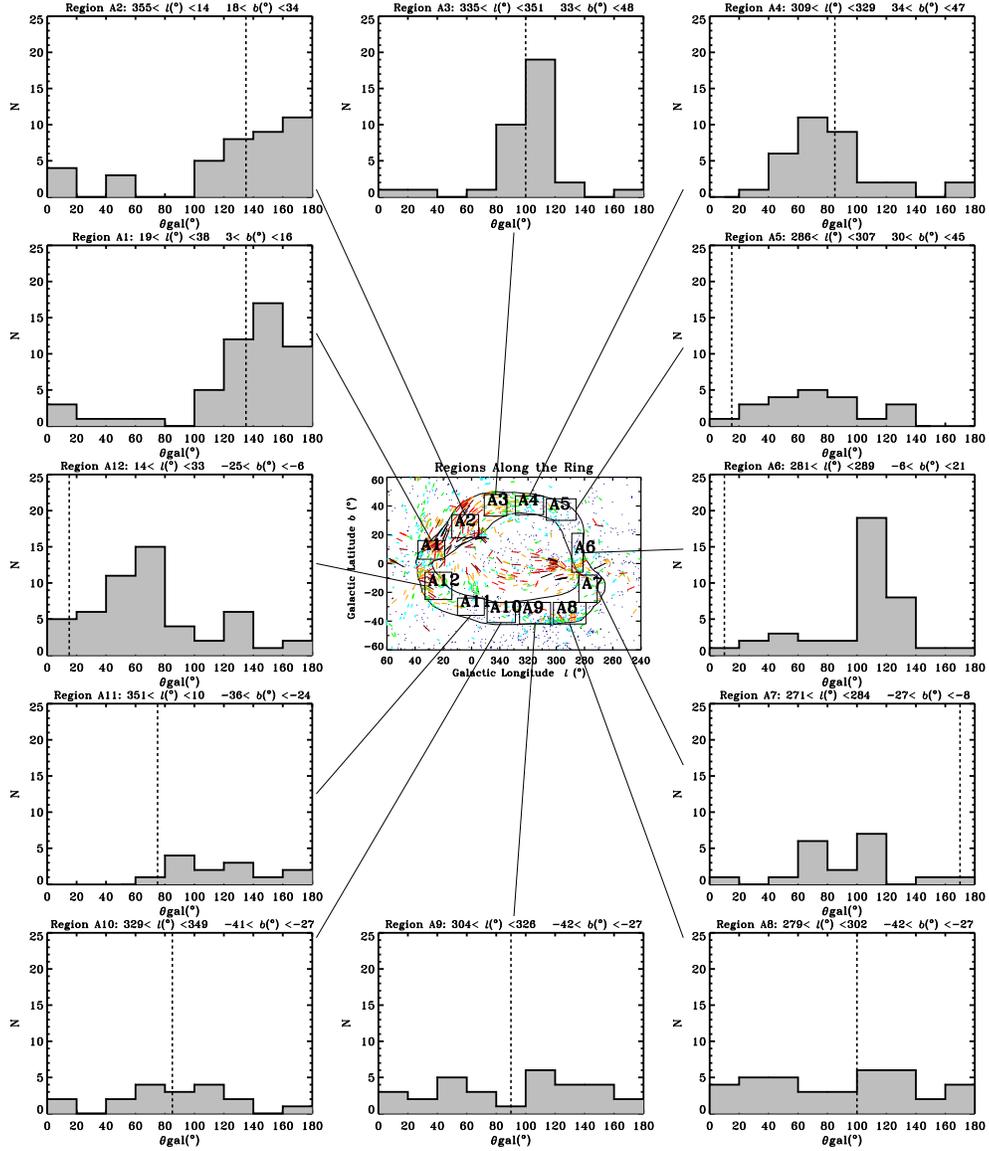}
      \caption{
               {\small
               Histograms of polarization vectors' direction to the same
               rectangular areas analyzed in Figure \ref{pdist_ring}.
               The dashed lines on the histograms roughly indicate the direction of
               the contour along each specific area.
               Polarization vectors are also plotted along the interface
               region at the center, as in Figure \ref{vecpol_integrated}.
               Note that left and northern areas (e.g., A1, A2, A3, and A4) show predominant
               vectors' orientation parallel to the ring contour (dashed lines), while at the right side
               and southern areas (e.g., A6, A7, A8, and A9) the main vectors' directions show no relation
               to the ring contour direction.
               }
              }
         \label{hist_ring}
   \end{figure}

\hspace{1em}
The analysis of the polarization vectors orientation for different
distance bands (Figure \ref{vecpol_intervals}), reveals that on the $180 - 210$ pc and $210 - 240$ pc diagrams
the polarization vectors along the right side direction 
do not present correlation with the ring contour orientation,
in contrast to the situation observed at the left side, where the vectors run
parallel to the ring contour. This relation between polarization vectors' mean direction
and the direction of the ring contour is further investigated on Figure \ref{hist_ring},
where we show histograms of the polarization angle ($\theta_{gal}$) to the same
previously studied rectangular areas. These histograms only account for the
polarized stars, i.e., with $P>0.1\%$ and $\Delta P/P < 0.2$. The dashed lines on
the histograms indicate the direction of the ring contour along each studied area.

In fact, the histograms associated to the left-northern areas (e.g., A1-A4) show a good
correlation between the ring contour direction and the main values of polarization
direction, as indicated by the coincidence of the histogram peak and the dashed 
line, that indicates the ring contour orientation.

\cite{iwan1980}, \cite{degeus1992}, and \cite{heiles1998} pointed out that
these lines-of-sight intercept several spherical ``shells" of interstellar material which
are centered on the sub-groups of Sco-Cen and are
most probably generated by stellar wind and SN explosion from these massive stars.
Furthermore, by analyzing the radio continuum maps ($408$MHz) by \citet{haslam1982}
and the relation between the shells' direction
and the polarization vectors, \citet{heiles1998} proposed that this configuration could be
created due to the distortion of the local magnetic field along constant ``longitude" lines
on the surface of the expanding shells.

In contrast, histograms related to the right side areas (e.g., A6 and A7) show that the
predominant polarization direction is mainly uncorrelated to the ring contour direction,
as indicated by the different position of the dashed line and the peak of the histogram.
Furthermore, the southern areas (A8, A9, and A10) show a large polarization
vectors dispersion, as suggested by the flat shape of the histogram. 

This could be related to the low polarization sky-projected
component of the local magnetic field (being viewed face-on)
or perhaps due to the presence of several interstellar sub-structures.
In fact, \citet{penprase1998} reported the existence of a diffuse interstellar filament
located in the direction of the A8 area at $d=230\pm30$ pc. A more detailed analysis
of the polarization vectors in the direction of this filament suggest a clear
relation with the morphology of such feature, with the polarization vectors
positioned perpendicular to the direction of the filament.

These results show that the behavior of the magnetic field direction
in relation to the direction of the ring contour is markedly different between
the left and right side regions.

\subsection{Implications on the models of the Local ISM}

\hspace{1em}
The existence (or not) of the interaction ring is closely related to the true
nature of the LC. Several models attempting to explain the origin of the LC
have been proposed, and have been divided here in three main classes, as discussed below.

On one side, the first class claims that the LC may have been created due to one or more SN explosions at the solar
neighborhood a few million years ago, which would
have swept the local neutral material and heated the interstellar gas up to $10^{6}$K
\citep{cox1974,cox1982,coxreynolds1987,gehrels1993,smith2001,maiz2001,berghofer2002,fuchs2006}.
The hot gas would have generated a local component on the all-sky observed soft X-ray emission (corresponding
to the ``Hot Local Bubble"). In fact, using
data from the ROSAT survey ($0.25$keV) and assuming a constant emissivity of the plasma,
\citet{snowden1990} and \citet{snowden1998} attempted to produce tridimensional maps of the
Hot Local Bubble.

On the other side, a second class of models suggest that the LC would be a part of
the nearby Loop I superbubble, in a sense that different star forming epochs from
the Sco-Cen OB association would have created different shock fronts, which expanded
asymmetrically in the direction of the Sun, i.e., a low density region between
the Galaxy's spiral arms \citep{frisch1981,frisch1983,frisch1995,wolleben2007}.
\citet{frisch2008,frisch2010} argues further that this model is consistent with the view that two
separate synchrotron emitting interstellar shells centered on different parts over the
Sco-Cen association can reproduce large-scale structures revealed by recent radio polarization surveys,
as proposed by \citet{wolleben2007}.
Furthermore, the existence of a global flux of interstellar material at the solar neighborhood
toward the opposite direction of Loop I \citep{lallement1995,lallement1998,cox2003},
i.e., toward $(l,b) \approx (186^{\circ},-16^{\circ})$,
supports the idea that some recent event may have caused the large-scale movement
of the interstellar material driven away from Sco-Cen.

A third class of models propose that the Local Cavity is not really a bubble in a sense that
its origin is not associated to SN explosions, being actually an ordinary cavity (the ``local void")
in the local interstellar medium \citep{bruhweiler1996,mebold1998}. The local cavity would be a low-density volume of space
defined by several interacting shells, like Loops I, II, III, and IV, the Gum Nebula, the Orion-Eridanus
Superbubble, among others.

If the interaction ring really exists, the first class of models provides a suitable scenario,
since the collisional models by \citet{yoshioka1990} would require
both bubbles (Local and Loop I) to be expanding, and in addition,
at least one of them should have reached the radiative stage of evolution prior
to the interaction in order to generate the proposed annular dense region,
as pointed out by \citet{egger_aschenbach1995}.

However, some issues have been recently presented regarding the idea that
the LC would have been created by such supernovae explosions. Specifically, recent studies revealed
that a major part of the observed soft X-ray emission could be related to a much closer
component, associated to the solar wind charge exchange effect at the heliosphere,
which have not been properly accounted on the previous analysis
\citep{cravens1997,freyberg1998,cox1998,cravens2000,robertson2001,robertson2003,lallement2004,welsh2009}.

Our data demonstrates that the classical concept of a unique bounding ring at the 
interaction area is misleading, in view of the largely discrepant distances
between the different parts of the annular area, and also due to the
widely different behavior of the magnetic field lines along the structure,
which suggest that the left and right sides of the proposed ring are possibly uncorrelated.
Figure 3 from the work by \citet{breit1996} shows a conceptual representation of the classical 
model attributed to the interaction ring.

Instead of the scenario proposed by \citet{egger_aschenbach1995},
our data suggest that this structure is quite distorted and fragmented,
unlike one would expect from a unique feature.
However, this evidence does not refute the first class of models, it simply
adds a new constraint to computational simulations which attempts to reproduce the characteristics
of the LC and Loop I through SN energy input.
Therefore, these models should account for the shape of the magnetic field lines, and 
reproduce the observed distortion along the interface area.

\begin{figure}[!b]
   \centering
   \includegraphics[width=0.5\textwidth]{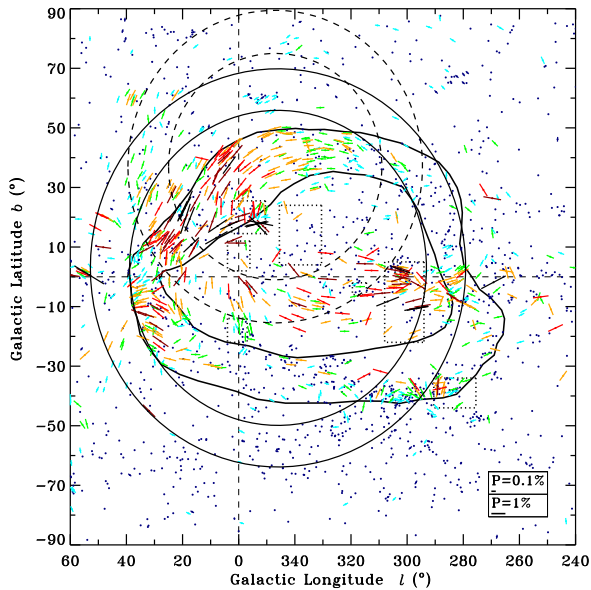}
      \caption{
              Distribution of polarimetric data in the direction of Loop I, used to illustrate
              the \citet{wolleben2007} model, represented by the projected positions of the S1 (solid line) and
              S2 (dashed line) shells. We have also included the contour of the ring proposed by \citet{egger_aschenbach1995},
              and the positions of several dark clouds located along the interface with Loop I.}
      \label{fig_wolleben}
   \end{figure}

On the other hand, the second and third types of model do not require the existence of an
interaction ring, since the occurrence of SN explosions inside the LC are not considered.
Particularly, the model provided by \citet{wolleben2007} allow the shape
of magnetic field lines to be tested by using polarimetric data,
similar to what has been accomplished by \citet{frisch2008,frisch2010}.
According to this scenery, Loop I is conceived as consisting of two synchrotron-emitting shells (namely S1 and S2),
that reproduces the large-scale magnetic field structure as revealed by polarimetric 
radio continuum surveys. As an illustration of this model, Figure \ref{fig_wolleben} shows 
the projected positions of both shells along with the optical polarimetric data distribution from 
this work.

From the point of view of our results, the advantage of such model is that it
doesn't require the existence of an interaction ring between the LC and Loop I.
However, it is important to point out that, referring to Figure \ref{fig_wolleben}, 
it is possibly misleading to perform a direct comparison
of the shells' {\it positions} with the vectors direction. Instead,
a valid analysis would demand a rigorous comparison between the polarization vectors and the
{\it summed projected magnetic field lines}, derived by \citet{wolleben2007},
along all distances and directions.
This means that the interweaving magnetic field lines related to the S1 and S2 shells 
should be projected in the plane-of-sky and thereafter vectorially summed
along each line-of-sight, similar to what has been discussed at the Appendix from \citet{wolleben2007}
(where the integrated Q and U Stokes parameters were computed based on the magnetic field structure).

Although there exist some correlations of the polarization vectors' direction with respect to the
shells' positions (especially toward E1 and E2 areas $-$ the North Polar Spur $-$
when compared to the left side of the S2 shell), such effect may be possible only when
the projected magnetic field follows the direction of the shell's border, which
is not true in general, considering the projections and the combined effect of both shells.
Therefore, it is soon to assert if this is an adequate model, and it would also be expected 
to reproduce the observed gas and dust column densities in the
lines-of-sight of the Loop I shells.

\section{Conclusions}
\label{conclusions}

We have carried out a polarimetric survey of $878$ Hipparcos stars in the general
direction of the Loop I superbubble's interface with the LC. Our sample was complemented with
the data from \citet{heiles2000} catalogue. The main results of the analysis
are summarized below:

\begin{itemize}

\item Along the ring structure proposed by \citet{egger_aschenbach1995},
the left side rise from $P\sim 0.2\%$ to $P\sim0.9-1.3\%$, which correspond to the ring's column density,
occurs at $\approx100$ pc, while at the right side regions the same
transition occurs only beyond $\approx250$ pc. This trend 
corroborates the color excess analysis by \citet{reis_corradi2008};

\item A gradual decrease in the polarization values is observed in the left-right direction
along the contour of the interaction ring;

\item The analysis of the polarization vectors direction along the interface revealed that
along the right side the predominant direction of the vectors do not present correlation with the
direction of the ring contour, in contrast to the situation observed at the left side,
where the vectors run parallel to the ring structure.

\end{itemize}

Altogether, these evidence confirm the distorted nature of the interstellar interface between the LC and Loop I.
The low polarization values toward some areas along the interaction ring suggest 
that, if it really exists, it is probably highly fragmented and twisted.
However, the configuration of the sky-projected component of magnetic field in relation to the direction of the ring
contour reveals markedly different behavior in the left and right sides.
This fact casts some doubt on the existence of the ring-like structure as a unique large-scale feature.

Our methods were mainly the analysis based on the distances to the interstellar structures
and the shape of magnetic field lines along the interface. These studies were used in
a qualitative comparison with the existent models for the local interstellar medium.
The existence (or not) of the ring is closely related to the true nature and associated origin models of the Local Cavity,
and shall deserve a particularly special attention.

\acknowledgements
The authors are grateful to the anonymous referee for the enlightening suggestions 
which greatly improved the paper.
The authors are also thankful to Fapemig (grant APQ 00154/08) and CNPq (131925/2007-5), which provided
financial support to the accomplishment of this research.
This work would not have been possible without the professionalism and competence
provided by the entire staff of LNA/OPD.



{\it Facilities:} \facility{LNA}.

\end{document}